\documentclass[10pt,journal,letterpaper,compsoc]{IEEEtran}
\usepackage[scaled=.92]{helvet}
\usepackage{times}
\usepackage{amsmath,amssymb,latexsym,float,epsfig,subfigure}
\usepackage{algorithm}
\usepackage{algorithmic}
\usepackage{graphicx}

\newcommand{\pd}{\partial}
\newcommand{\D}{\nabla}

\newcommand{\sphere}{\mathcal{S}^{2}}

\usepackage{parskip}
\begin{document}
\title{Volumetric Mapping of Genus Zero Objects \\via Mass Preservation}

\author{Romeil~Sandhu*,~\IEEEmembership{Member,~IEEE,}
       Ayelet~Dominitz*,~\IEEEmembership{Member,~IEEE,}
        Yi~Gao,~\IEEEmembership{Member,~IEEE,}
        and Allen Tannenbaum,~\IEEEmembership{Fellow,IEEE}
\IEEEcompsocitemizethanks{\IEEEcompsocthanksitem R. Sandhu is with Harper Laboratories, LLC, Atlanta GA, 30332.  A. Dominitz is with Elbit Systems Ltd, Haifa Israel. Y. Gao is with the Brigham and Women's Hospital, Harvard Medical School, Boston, MA 02115. A. Tannenbaum are with the School
of Electrical and Computer Engineering, Boston University, Boston,
MA, 02215.\protect\\
E-mails: {romeil.sandhu@harperlabs.com, ayelet.domash@gmail.com, tannenba@bu.edu}}}
\IEEEcompsoctitleabstractindextext{%
\begin{abstract}
In this work, we present a technique to map any genus zero solid object onto a hexahedral decomposition of a solid cube. This problem appears in many applications ranging from finite element methods to visual tracking.  From this, one can then hopefully utilize the proposed technique for shape analysis, registration, as well as other related computer graphics tasks.  More importantly, given that we seek to establish a one-to-one correspondence of an input volume to that of a solid cube, our algorithm can naturally generate a quality hexahedral mesh as an output.  In addition, we constrain the mapping itself to be volume preserving allowing for the possibility of further mesh simplification.  We demonstrate our method both qualitatively and quantitatively on various 3D solid models.
\end{abstract}

\begin{keywords}
Hexahedral Mesh, Mesh Generation, Conformal Mapping, Mass Preservation
\end{keywords}}

\maketitle

\IEEEdisplaynotcompsoctitleabstractindextext
\IEEEpeerreviewmaketitle

\section{Introduction}
With the ever increasing complexity of 3D digital models, the fundamental need of simplifying the task of data analysis has arisen as an important yet challenging problem.  For example, given a volumetric object, one would like to properly define a metric capable of quantifying (dis)similarities.  This has numerous applications ranging from shape retrieval to registration in both computer graphics and the general vision community \cite{haker04}.  In this note, we extend our previous work of texture mapping \cite{ayelet10} by suggesting a volume preserving mapping capable of transporting a given solid object onto a solid cube.  Consequently, qualitative and quantitative analysis tasks can then be performed on the cube's  Euclidean coordinate system as opposed to an arbitrary three dimensional shape.  Moreover, given that our ``target'' is a solid cube, the algorithm is able to naturally employ a hexahedral mesh structure.  This is highly desirable for finite element analysis and physically-based deformations or simulations.  The algorithm itself can be decomposed into three main steps:
\begin{enumerate}
\item Construct the intermediate surfaces and subdivisions of a volumetric object if a surface mesh is given.
\item Construct an initial diffeomorphic volumetric map that transports the solid object onto the solid cube.
\item Improve the initial volumetric map such that it preserves volume (no loss of ``information'').
\end{enumerate}

\textbf{Key Contributions:}  We develop an efficient algorithm that is able to robustly and automatically compute a volume preserving mapping from any solid model to the solid cube.  Aside from generating a quality hexahedral mesh, the cube allows one to transfer the task of feature analysis onto a Euclidean domain. This is particularly important in longitudinal medical studies in which a patient is examined over a period of time.  From this, the goal might be to uncover subtleties with respect to a certain medical structure being imaged.  In particular, a common hypothesis of being able to provide early detection of Schizophrenia might be linked to shape abnormalities in the the corpus callosum \cite{mohan10}.  By providing a canonical domain such as the cube, it is our belief that the proposed algorithm will aid in such longitudinal studies.

Moreover, by invoking the constraint of volume preservation, one can then transfer volumetric functions such as color, material, solid texture, as well as strain/stress tensors to and from the original model of interest to the cube without jeopardizing the ``information'' itself.  For example, a flattened representation of colon surface is helpful for the detection of colon polyps \cite{haker00} and a flattened representation of vessel surface is useful for the study of the correlation between wall shear stress and the development of atherosclerosis \cite{zhu05}.  While the motivation for these examples is to preserve area and minimize distortion, one can equally extend the thought process to minimizing volume distortion.  Thus, the constraint is utilized so that it will allow for the user to be able to compare volumetric functions on the cube as if it were on the initial object with minimal distortion.

The remainder of this paper is organized as follows: In the next section, we briefly revisit some of the key results that have been made pertaining to both volumetric mapping and meshing algorithms. We then discuss the proposed algorithm with the appropriate details in Section~\ref{sec:overview_theory}. Section~\ref{sec:NumericDetailsImp} provides additional details regarding numerical implementation. In Section~\ref{sec:mesh_experiments}, we present experimental results on various 3D solid models both from a qualitative and quantitative perspective. Finally, we discuss future work in Section~\ref{sec:mesh_future_work}.
\begin{figure*}
\begin{center}
\includegraphics[width=5in,height=1.65in]{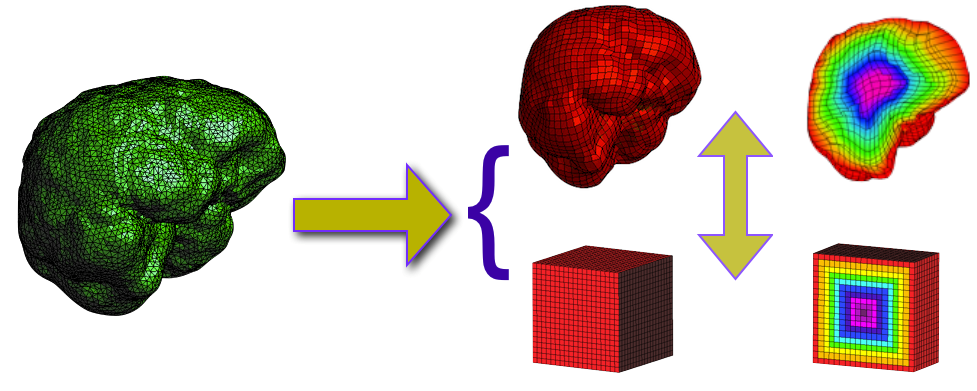}
\end{center}
\label{fig:sweetspot_mesh}
\caption{Given an input such as a triangulated surface mesh, we propose an algorithm capable of constructing a volume preserving mapping from a solid model to a solid cube.  In addition, mapping a solid model to a cube allows for us to naturally (and simultaneously) generate a regular hexahedral mesh as opposed to tetrahedral meshes.}
\end{figure*}

\section{Related Work}
\label{sec:related_work}
In what follows, we detail methodologies that are related to the approach proposed in this work.
\subsection{Volumetric Mapping Algorithms}
Volumetric mapping algorithms aim at establishing a one-to-one bijective correspondence between a model of interest to that of another model, typically represented by a simpler volume (e.g., a cube or a sphere).  Previous work related to our methodology has concentrated on finding harmonic volumetric maps from a solid model to the sphere using tetrahedral meshes \cite{wang04,ju05,li09}. These works employ harmonicity, defined by the vanishing Laplacian, to represent the smoothness of their mapping function. In particular, Wang \emph{et al.} \cite{wang04} proposed a variational approach whereby they define a harmonic energy on a tetrahedral mesh to compute the necessary discrete volumetric harmonic maps. Specifically, they begin by conformally mapping the boundary of the 3D volume to a sphere, and then minimize the proposed harmonic energy while keeping the surface fixed.  Ju \emph{et al.} \cite{ju05} offer a solution whereby the authors generalized the idea of mean value coordinates \cite{floater03} for surfaces to volumes. Recently, Li \emph{et al }\cite{li09} introduced a boundary method denoted as the method of fundamental solution (MFS). That is, they construct the harmonic volumetric mapping through a meshless procedure by using only the boundary information.  In contrast to these techniques, the proposed algorithm in this note differs in three main ways.  Firstly, we attempt to construct a canonical domain in euclidean space whereby we wish to derive a volumetric map that transports a solid model onto a cube (as opposed to a sphere).  Secondly, the output of our algorithm is not a tetrahedral mesh, but a quality hexahedral mesh structure.  Lastly, while these methods attempt to minimize a harmonic energy with the constraint that the mapping itself is smooth, we aim at developing a volume preserving mapping.

In addition to the above methods, extensive work has been done using other domains such as polycubes \cite{he09,xia101,han10,li10}.  While the work presented in this paper focuses on the solid cube, different domains may also be used.  Specifically, the authors in \cite{he09} successfully employed an efficient and novel methodology to construct polycube maps for surfaces with complicated topology.  At the same time, the method requires little to no user interaction.  Recent work by \cite{xia101} provides a solution in which the authors partition or decompose the volumetric object as a direct product of a two dimensional surface and one dimensional curve.  Generalizing the polycube representation, \cite{li10} develops a trivariate hierarchal spline scheme to properly represent volumetric data.  More recently, \cite{gregson11} presents a technique for computing low-distortion volumetric polycube deformations of generals shapes and resulting hexahedral meshes.

\subsection{Meshing Algorithms}
In this section, we review related works focusing on hexahedral meshing algorithms with the understanding that there exists a wide variety of possible approaches to construct volumetric meshes \cite{zhang10,nieser11}. Generally, the hexahedral mesh generation algorithms are categorized by two classes: structured and unstructured methods. Strictly speaking, a structured mesh is characterized by all of the interior mesh nodes having an equal number of adjacent elements. On the other hand, unstructured meshes relax the nodal valence requirement, allowing any number of elements to meet at a single node. We note that the mesh we construct falls into the structured mesh category, which includes mapping techniques \cite{cook82} and submapping approaches \cite{white95}.  In addition to these frameworks, structured meshing algorithms have arisen in the form of octree approaches \cite{schneiders97,zhang06}, multiblock methods \cite{dannenhoffer91,grosland09}, and sweeping algorithms \cite{knupp98,roca04,shepherd00}.  While it is beyond the scope of this note to detail all of these methods (and is by no means exhaustive), we refer the interested reader to a survey conducted by Owen \cite{owen98}.  In what follows, we will further constrain our discussion to approaches related to mapping techniques.

For mapped meshing to be applicable, the volumetric objects of both the model of interest and the ``destination'' must have equal numbers of divisions with similar topology.  However,  this can often be impossible for an arbitrary geometric configuration or can involve considerable user interaction to decompose geometry into mapped meshable regions.  In order to reduce human interaction, work has be done in recent years through the CUBIT \cite{cubit} project at Sandia National Labs to automatically recognize features \cite{tautges97} and decompose geometry \cite{liu96,mitchell00} into separate mapped meshable areas and volumes.

In a similar fashion, methods have been devoted to what is known as sub-mapping \cite{white95}. In contrast to above works, which decompose the geometry directly, the method proposed by White \emph{et al.} \cite{white95} determines an appropriate virtual decomposition based on corner angles and edge directions. These (mapped-)meshable regions are then meshed in a separate manner. Although this approach is suitable for shapes and volumes that have well defined corners and cube-like regions, it is hindered from its seemingly inflexibility to incorporate additional geometric topologies to the predefined topology. Consequently, hexahedral meshing for these particular cases result in extensive decomposition of the solid into the predefined topologies by massaging the elements to fit the geometry at hand.

In addition to these related works, Martin and Cohen \cite{martin10} recently presented a methodology for hexahedral meshing that is able to handle higher order genus species through B-splines and T-splines.  In contrast to these proposed approaches, our algorithm only requires the object to be of genus zero topology. By {\em genus zero} we mean that the solid has no holes (i.e., the surface is contractible).  However, if the object is not of genus zero, it may be processed by separating the object into several genus zero parts.

\section{Proposed Algorithm and Theory Overview}
In this section, we provide the details of the proposed algorithm.  In particular, we formulate the problem of interest and then proceed to discuss each of the major steps to compute a volume preserving mapping. 
\label{sec:overview_theory}
\subsection{Problem Statement:  A Volume Preservation Map}
We first give a precise mathematical formulation of our problem. We wish to compute a volumetric map $f:\mathbb{R}^{3}\mapsto{R}^{3}$ from a solid cube $C^3$ to a given compact volume $M$, which is equivalent to building a one-to-one correspondence between them. We define a positive density function over our domains $C^3$ and $M$, denoted by $\mu_C^3$ and $\mu_M$, respectively. We assume that the total mass associated with each of the volumes is equal:
\begin{equation}
\int_{C^3}\mu_{C^3}\left(x\right)dx=\int_{M}\mu_{M}\left(y \right)dy
\end{equation}
where $dx$ and $dy$ are the standard volume forms on $C^3$ and $M,$  respectively. If this assumption is not satisfied, we can always scale one of the density distributions to make the total amount of mass equal.

We constrain our mapping to be mass preserving, so that it will satisfy \begin{equation} \mu_{C^3}=\vert\nabla f\vert\mu_{M}\circ f . \label{Eq: Jacobian}\end{equation} Here $|\nabla f|$ denotes the determinant of the Jacobian map $\nabla f$ and $\circ$ represents composition of functions. This equation is often referred to as the \textit{Jacobian equation}, which constrains the mapping to be mass preserving (MP) with respect to the given density functions. A mapping $f$ that satisfies this property may thus be thought of as defining a redistribution of a mass of material from one distribution $\left(C^3,\mu_{C^3}\right)$ to another distribution $\left(M,\mu_{M}\right).$ Assuming that the total volumes of $C^3$ and $M$ are equal, we can then say that a diffeomorphism is {\em volume preserving}, if it maps the volume form of $C^3$ to the volume form of $M$.  However, in order to compute this mapping and the corresponding hexahedral mesh, we will also need to define the resolution of the volumetric object $M$. This is discussed next.
\subsection{Constructing the Intermediate Subdivisions}
For any given meshing algorithm, there is an appropriate resolution parameter that can be altered so that the output mesh is of the proper detail.  In our work, we denote $N$ to be this resolution parameter and we note that is the only parameter required as an input.  Moreover, this parameter dictates the number of confined surfaces that we use for the mapping and relates the size of the ``destination'' cube that we would like to transport our model onto (e.g., the more surfaces results in a higher resolution).
 \begin{figure}[t]
 \begin{center}
 \includegraphics[width=3in,height=1in]{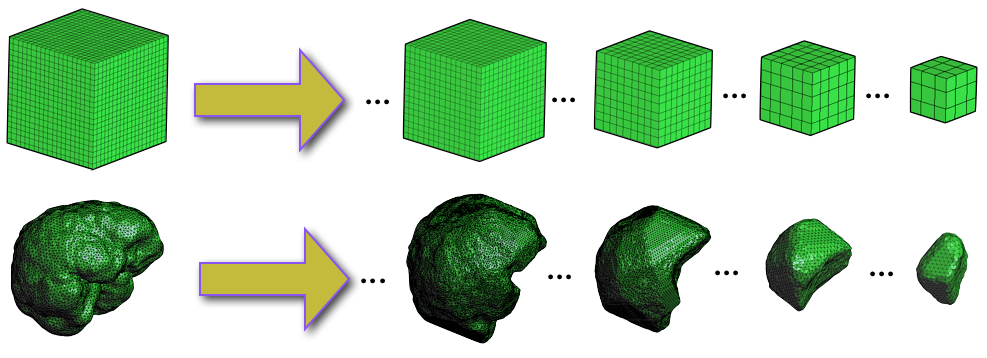}
 \end{center}
 \caption{An illustration of the decomposition step during the formation of the initial map.  The cube and model are decomposed into $N$ shells, where $N$ is a user defined parameter.  Note:  The size of grid sizes do not change.  This appearance is a result of the visualization.}
 \label{fig:decomposition_mesh}
 \end{figure}
We first form a cube volume $C^{3}$ from small $2N \times 2N \times 2N$ cubes. This gives us a hexahedral mesh of cubic volume. We then partition the solid volumetric cube into N confining cubical surfaces \{$C^{2}_{k} ; k = 1 ,. . . , N$\}.  Note that the denotation of ``confining cubes'' represents the shells or surfaces of the volumetric cube as seen in Figure \ref{fig:decomposition_mesh}.

Specifically, we generate these subdivisions of the cube by iteratively eroding the outer layer of our solid cube and taking the boundary of our new cube as the next confining surface. This is shown in the top row of Figure \ref{fig:decomposition_mesh}, where a few of these confining cubic surfaces are displayed.  Note, the grid or element size does not change as viewed in the figure.  This appearance in the size change of elements is due to the effect of visualization. With this being said, the goal is to map each of these cubic surfaces to their respective confining surfaces that lie within the original object.

In order to construct the intermediate surfaces as shown in the bottom row of Figure \ref{fig:decomposition_mesh}, let us start with a binary representation $\beta$ of the input object $I_{N}$.  In other words, $\beta$ is labeled ``1'' on the surface of $I_{N}$ and region contained within the surface of $I_{N}$; otherwise, it is labeled ``0''.  While erosion works to construct the intermediate surfaces of our cube, this method will fail on the input object due to varying topology. Thus, in order to generate the intermediate subdivisions, we perform classical segmentation via the GAC framework \cite{caselles97}.

\begin{figure}[t]
 \begin{center}
 \includegraphics[height=1.75in]{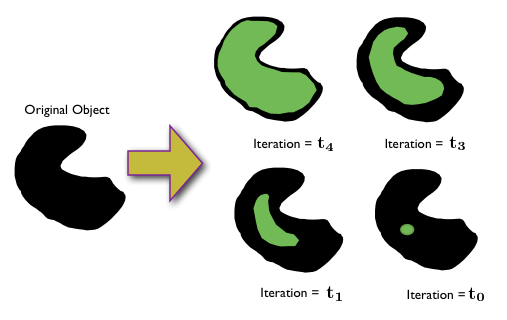}
 \end{center}
 \caption{An illustration of 2D Geometric Active Segmentation during evolution.  Object of interest is represented by the color black.  The result of the segmentation at a given time is represented by the color green.  Subdivisions or confining surfaces are generated by ``pulling out'' these intermediate results.}
 \label{fig:active_con}
 \end{figure}

Of course, if we are given the surface of the object and its binary representation, then the problem has been solved from a segmentation viewpoint.  However, our interest lies in the evolution of the segmentation when given an initial surface.  That is, by using a segmentation method such as active contours, we will ``pull out'' $N-1$ surfaces over time until the algorithm segments the original surface.  For example, Figure \ref{fig:active_con} shows a typical evolution of 2D GAC segmentation of a binary object taking place and the resulting segmentation results at different times.  The corresponding 2D segmentation results are analogous to the 3D confining surfaces shown at the bottom of Figure \ref{fig:decomposition_mesh}.  Moreover, with the use of coupled active contours \cite{yezzi02}, one can robustly generate subdivisions without overlap.  In a similar fashion, Xia et. al. \cite{xia10} have alternatively used level-sets via Green's functions to parametrize volumetric objects.   Nevertheless, for the sake of clarity, we opt to utilize the classical GAC framework in this current work and also discuss its inherent disadvantages in Section \ref{sec:NumericDetailsImp}.

Thus, we consider the problem of segmenting $\beta\in\mathbb{R}^{3}$.  To this end, we enclose a surface $\mathcal{S}$, represented as the zero-level set of a signed distance function $\phi:  \Re^{3}\rightarrow\Re$, such that $\phi < 0$ represents the inside of $\mathcal{S}$ and $\phi > 0$ represents the outside of $\mathcal{S}$ \cite{osher2003}.  Our goal is to evolve the surface $\mathcal{S}$, or equivalently $\phi$, so that the surface $\mathcal{S}$ would match the original surface $I_{N}$.  From an optimization standpoint, we seek to minimize a cost functional of the general form $E = \int_{\gamma}\psi(x,t)dx$ over a family of surfaces.  Specifically, we let $\psi(x,t)$ be the classic Chan-Vese energy \cite{chanandVese2001}:
 \begin{align}
E = \int_{\Omega}H_{\epsilon}(\phi)& (\beta(x)-\mu_{\text{in}})^{2}\\ \notag
         + &(1-H_{\epsilon}(\phi))(\beta(x) - \mu_{\text{out}})^{2}  dx
\end{align}
where $H_{\epsilon}(.)$ denotes the Heaviside functional as
\begin{equation}
 H_{\varepsilon}(\phi) = \left\{ \begin{array}{rl}
  1 &\mbox{$\phi<\epsilon$} \\
  0 &\mbox{$\phi>\epsilon$} \\
  \frac{1}{2}(1 + \frac{\phi}{\epsilon} + \frac{1}{\pi}\sin(\frac{\pi \phi}{\epsilon})) &\mbox{otherwise } \phi
    \end{array} \right.
\end{equation}
and $\mu_{\text{in}}$ and $\mu_{\text{out}}$ are given as
\begin{equation}
\mu_{\text{in}} =  \frac{\int_{\Omega}H_{\varepsilon}(\phi)\beta(x)dx}{\int_{\Omega}H_{\varepsilon}(\phi)dx} \quad\quad
\mu_{\text{out}} =  \frac{\int_{\Omega}(1-H_{\varepsilon}(\phi))\beta(x)dx}{\int_{\Omega}(1-H_{\varepsilon}(\phi))dx}.\notag
\end{equation}In particular, minimizing the above energy functional will allow the curve to move in the direction such that the mean inside and outside the curve, $\mu_{\text{in}}$ and $\mu_{\text{out}}$, is maximally separate in the $L_{2}$ sense.  As a result, we are able to generate $N$ intermediate surfaces $\{I_{k}; k = 1, . . ., N\}$ confined to lie inside or on the interested volumetric object.  Note that $M=I_{1}\cup I_{2},...,\cup\text{ } I_{N}$. We are now ready to begin to construct our initial mapping between the cube and our model.

\subsection{Constructing the Initial Map}
\label{sec:initial_map}
 Utilizing the conformal mapping technique proposed by Angenent \emph{et al.} \cite{angenent02}, we derive
 the connectivity and the corresponding map between the solid model and the solid cube.  Like that of the cube,
 which provides a canonical domain for volumetric objects, the sphere yields a simplified domain for surfaces.
From this, our goal is to employ the conformal mapping technique in
order to map \emph{each} of the intermediate surfaces onto the
sphere $S^{2}$, perform interpolation, and take the inverse mapping
to generate an initial diffeomorphism between the cube and model.
In what follows, we will focus on conformally mapping intermediate
surfaces $\{I_{k}; k = 1, . . ., N\}$ of our solid model $M$ with
the understanding that we also perform the same procedure for the
cubic surfaces \{$C^{2}_{k} ; k = 1 ,. . . , N$\}.

In doing so, we first remove a triangle $p$ of the specified mesh $I_{k}$, and then solve the Dirichlet problem
to obtain the conformal map $u_{k}:I_{k}\text{\textbackslash} \{p\}\mapsto S^{2}\text{\textbackslash}\{\text{north pole}\}$.
This shown in Figure~\ref{fig:initial_map} for both of the confined surfaces defined previously.
Solving the Dirichlet problem amounts to conformally mapping the altered surface onto the complex plane.
This then allows us to use the inverse stereo projection to project the plane onto the unit sphere.
The boundary $p$ is mapped to a triangle around the ``North Pole'' of $S^{2}$.

More precisely, we start with a manifold represented by a triangular mesh with $W$ vertices,
from which we remove one triangle $\triangle ABC$, and then perform the following steps:
\begin{itemize}
\item Calculate the matrix $D$:

$D$ is a sparse and symmetric $W \times W$ matrix whose non-zero elements $D_{PQ} (P \neq Q)$ are:
\[D_{PQ}=-\frac{1}{2}\left(\cot \angle R + \cot \angle S \right) \] where $\angle R$ is the angle at the vertex $R$
in the triangle $\triangle PQR$ and $\angle S$ is the angle at the vertex $S$ in the triangle $\triangle PQS$.

\begin{figure}[h]
    \begin{center}
   \includegraphics[scale=0.15]{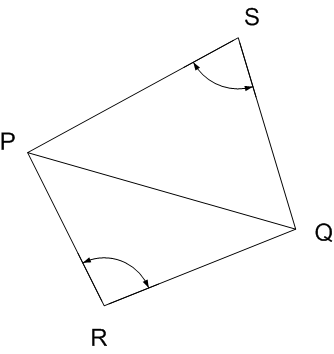}
    \end{center}
\label{fig:conformal_angles1}
\end{figure}

\begin{figure}[h]
    \begin{center}
   \includegraphics[scale=0.15]{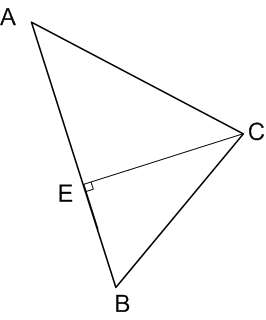}
    \end{center}
\label{fig:conformal_angles2}
\end{figure}
Notice, that $D_{PQ} \neq 0$ only if $P$ and $Q$ are connected by some edge in the
triangulation.
The diagonal elements $D_{PP}$ satisfy: \[\sum_P D_{PQ} = 0.\]
\item Calculate the vectors $a$ and $b$:

$a$ and $b$ are sparse $W$ vectors with an entry for each vertex. Their non-zero elements are found
at the vertices ${A,B,C}$ of the triangle $\triangle ABC$ that we removed from the mesh:

\begin{equation}a = \begin{cases} 0 & Q\notin \triangle ABC \\
                \frac{-1}{\lVert B-A \rVert}& Q = A \\
                \frac{1}{\lVert B-A \rVert}& Q = B \\
                0 & Q = C \\
                \end{cases}\notag \end{equation}

\begin{equation}b = \begin{cases} 0 & Q\notin \triangle ABC \\
                \frac{1-\theta}{\lVert C-E \rVert}& Q = A \\
                \frac{\theta}{\lVert C-E \rVert}& Q = B \\
                \frac{1}{\lVert C-E \rVert}& Q = C \\
                \end{cases}\notag \end{equation}
where $E$ is the orthogonal projection of $C$ on $AB$ and \[\theta = \frac{\left< B-A,C-A\right>}{\lVert B-A \rVert ^2}.\]

\item Solve the linear system of equations :
\begin{equation}\begin{cases} Dx = a \\ Dy = b\end{cases}.\notag \end{equation}

\item Map the $x$ and $y$ coordinates from the plane to the unit sphere using inverse stereographic projection:
\[ x + iy \rightarrow \left(\frac{2x}{1+r^2},\frac{2y}{1+r^2},\frac{2r^2}{1+r^2}-1\right).\]

\end{itemize}

\begin{figure}[!t]
\begin{center}
\includegraphics[width=2.5in]{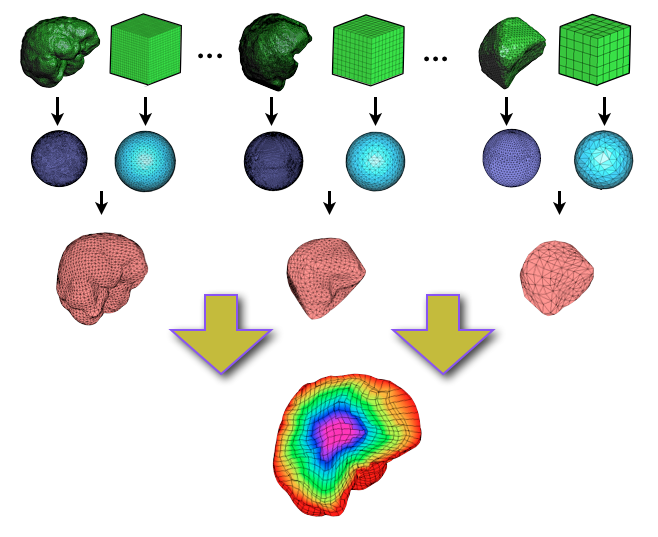}
\end{center}
\caption{An illustration of conformal mapping step during the formation of the initial map.
The intermediate surfaces of the cube and model are mapped to the sphere, interpolation is performed, and an inverse mapping is performed to generate initial map.}
\label{fig:initial_map}
\end{figure}

\begin{algorithm}
\caption{Create\_Initial\_Map(MODEL, N)}
\label{Alg: step1}
\begin{algorithmic}
\STATE $\backslash\backslash$ 1)  Initialize Parameters
\STATE $N \quad \Leftarrow$  Resolution parameter
\STATE  $L \hspace{2pt}\quad \Leftarrow $ Steps required for GAC to converge
\STATE $dt \hspace{9pt}\Leftarrow L/N$
\STATE $C^3 \hspace{5pt} \Leftarrow 2N\times 2N\times 2N$ regular hexahedral mesh
\STATE
\STATE $\backslash\backslash$ 2) Set Boundary and Map
\vspace{3pt}
\STATE $ C^2\{N\} \Leftarrow boundary(C^3) $
\STATE $S\{N\}\hspace{5pt} \Leftarrow boundary(\text{MODEL})$
\STATE $ C^3 \hspace{18pt} \Leftarrow erode\_layer(C^3)$
\STATE
\STATE $\backslash\backslash$3) Create Intermediate Surfaces and Map
\FOR{$i=1$ to $N$}
\IF{i $>$ 1}
\STATE $ C^2\{i\} \Leftarrow boundary(C^3) $
\STATE $ C^3 \hspace{13pt} \Leftarrow erode\_layer(C^3)$
\STATE $ V\{i\}\hspace{4pt} \Leftarrow \text{Run GAC (MODEL}, L - i\cdot\text{ step)}$
\STATE $ S\{i\}\hspace{5pt}  \Leftarrow boundary(V\{i\})$
\ENDIF
\STATE $ \hat{S}\{i\} \hspace{35pt}\Leftarrow map(S\{i\})$
\STATE $ \hat{C}^2\{i\} \hspace{30pt}\Leftarrow map(C^2\{i\})$
\STATE $ \bar{S}^{-1}\{i\}\hspace{26pt} \Leftarrow interpolate(\hat{S}\{i\}^{-1})$
\STATE $Remesh\_S\{i\} \Leftarrow $ $\bar{S}^{-1}\{i\}\circ(\hat{C}^2\{i\})$
\ENDFOR
\STATE $ connenctivity \hspace{8pt}\Leftarrow \mbox{cube connenctivity}$

\end{algorithmic}

\end{algorithm}
To clearly illustrate our scheme, we have provided pseudo code above.  Moreover, one can also view the
above method as the cotangent discretization of the well known Laplace-Beltrami operator.
Also, although the conformal mapping has the property that it maintains the local geometry of the original surface,
it distorts the area of the mapped surface.  As a result, we need to ensure that initial mapping is proper
quality since portions of the object where sharp features exists may be unfairly treated.  This is discussed next.

\subsubsection{Ensuring a Quality Initial Map}
We require the initial mapping to be of quality, in the sense that no volumes are mapped to infinitesimally small volumes.
In case of very smooth surfaces it is sufficient to use the conformal mapping solely. However, in cases of very undulated surfaces, the algorithm discussed above may map areas with high curvature to very small areas. Accordingly the corresponding hexahedral volumes will be mapped to very small volumes.  Recent work focusing on the problem of small area distortion during the construction of the conformal map is discussed by Afalo and Kimmel \cite{kimmel2011}. In this section, we address this issue via area preservation.

To this end, we quantify the change of area in the mapping $u_k$ as the density function $\mu_k$ at each point on the sphere $\sphere$, so that the integral on the flattened surface $\int_{\sphere}\mu_k dx$ will give us the area measure of the original surface. In continuous settings this density function is the determinant of the Jacobian of $u_{k}^{-1}$:\[ \mu_{k}=\vert\D u_{k}^{-1}\vert.\] In the discrete settings we calculate the area distortion for each surface element  as the ratio of the area of the surface element on the original surface, $C^{2}_{k}$ or $I_{k}$, to the area of its corresponding surface element on the sphere $\sphere$.

For each surface, we use the method proposed by Moser \cite{moser65} to ensure a quality map.  Specifically, we solve two differential equations on the sphere as follows.  First solve: \[\Delta\Theta=1-\mu_{k}\left(x\right)\] for $\Theta$, with \[ \mu_{k}=\vert\D u_{k}^{-1}\vert\] and $u_{k}$ as our conformal mapping for an intermediate surface. We then calculate the vector field $Y_t$ as \[ Y_{t}=\frac{-\D\Theta}{\left(1-t\right)\mu_k+t},\] for $t = 0\ldots1$. Finally we integrate the vector fields $Y_t$ to compute $w=w_1:$
\[\frac{d w_t(x)}{dt} = Y_t(w_t(x)).\]

This process is performed for each intermediate surface. The composition of this mapping $w_k$ and the conformal mapping $u_k$ is our area preserving mapping from the intermediate surfaces to the sphere. We will denote this mapping $q_k$: \[q_k = w_k\circ u_k.\]
\begin{figure}[!t]
\mbox{ \centering
\parbox{.45\columnwidth}{
   \quad\quad\quad  \includegraphics[width=.9in]{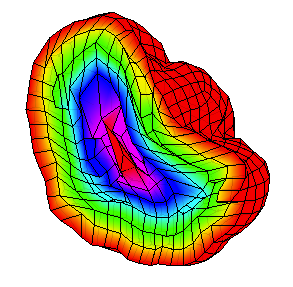}
      }\quad\quad\hspace{3pt}
\parbox{.45\columnwidth}{
    \includegraphics[width=.8in]{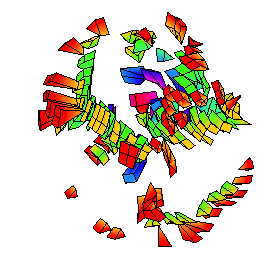}
      }}
    \mbox{ \centering
    \parbox{.45\columnwidth}{\centering
    (a)}
    \parbox{.45\columnwidth}{\centering
    (b)}}\\
\mbox{ \centering
\parbox{.45\columnwidth}{
  \quad\quad\quad\includegraphics[width=.8in]{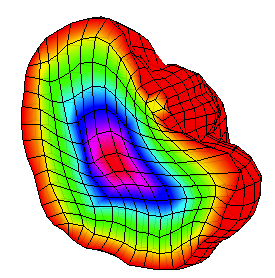}
      }\quad\quad\quad\hspace{3pt}
\parbox{.45\columnwidth}{
    \includegraphics[width=.7in]{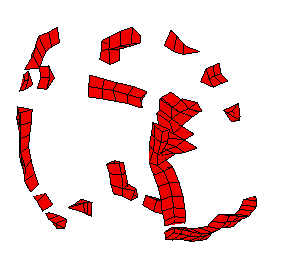}
      }}
    \mbox{ \centering
    \parbox{.45\columnwidth}{\centering
    (c)}
    \parbox{.45\columnwidth}{\centering
    (d)}}
\caption{The effect of mesh smoothing and volume preservation with respect to concave hexahedron.  (a) Initial map formed after construction of intermediate surfaces. (b) Concave elements from the initial map.(c) Final map with smoothing and evolving the mesh with our volumetric flow.  (d)  Concave elements from the final map.}
\label{fig:concave_smooth}
\end{figure}

Of course, we are interested in the connectivity of the solid model to the cube, and this involves the inverse mapping $q_{k}^{-1}$ with interpolation.  However, before doing so, we perform the same procedure above on the cubic surfaces  \{$C^{2}_{k} ; k = 1 ,. . . , N$\} to yield a conformal map $Q_k$.  Then we interpolate the inverse mapping of $q_k$ on the sphere, denoted as $\tilde{q}_k$, and map the nodes of the cubic surfaces $C^2_k$ to the intermediate surfaces with $\tilde{q}_k^{-1}\circ Q_k$. Now we have reconstructed the intermediate surfaces using the nodes (and connectivity) of the cubic surfaces. We can connect between the surfaces using the connectivity of the cube to obtain a meshing of the given solid that corresponds to the mesh of the solid cube.  More importantly, we now have an initial volumetric mapping defined earlier to be $f=\tilde{q}_k^{-1}\circ Q_k$.
\subsection{Improving the Initial Map}
Up until now, the map that has been generated is diffeomorphic, but has not been constrained to preserve volume.  In other words, the previous conformal mapping process for each of the confining surfaces of the cube and of the object itself was done in order to preserve area and remove distortion.  Now that we have derived the connectivity between the cube and the interested object, we are able concatenate the surfaces to form an initial hexahedral mesh of our original object. From this, we need to evolve our hexahedral mesh in a manner that satisfies the volume (not area) constraint. Since the mapping originates from a regular cube, this amounts to having a mesh with its elements having equal volume.  However, as with any method, there will be deviations and we capture those deviations through a quantitative evaluation of volume variance as demonstrated in Section \ref{sec:mesh_experiments}.

At any rate, we attempt to preserve mass by adjusting the method
proposed by Moser and Dacorogna \cite{moser65,dacorogna90} for the
volumetric setting.  In addition to just presenting our volumetric
flow, we outline a proof that guarantees the existence of a volume
preserving diffeomorphism (assuming of course the two solids have
the same volume).  We refer the reader to \cite{moser65} for
complete details and as an aid with regards to the information
presented next.

\subsubsection{Existence of Volume-Preserving Flows}
With a slight abuse of notation, we let $M$ and $N$ be diffeomorphic
compact solids with the same total volume. Accordingly, we denote
$f:N\to M$ be a diffeomorphism and $\omega_1$ be the pullback of the
volume-form of $M$ under $f$ and let $\omega_0$ be the volume form
of $N$ itself. The $\omega_i$ are three-forms with the same integral
over $N$, and we want a diffeomorphism $g:N\to N$ with
$g^*(\omega_1) = \omega_0$. Given $g$, the volume preserving map of
$N$ onto $M$ is $f\circ g$.
\begin{figure}[!t]
\mbox{ \centering
\parbox{.45\columnwidth}{
     \quad\quad\includegraphics[width=1.1in]{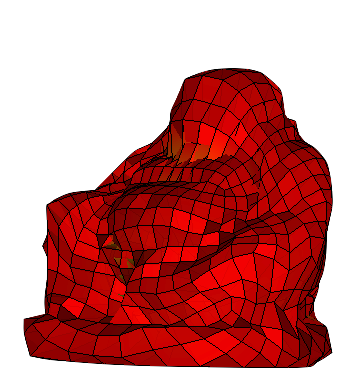}
      }\quad\quad
\parbox{.45\columnwidth}{
   \includegraphics[width=1.1in]{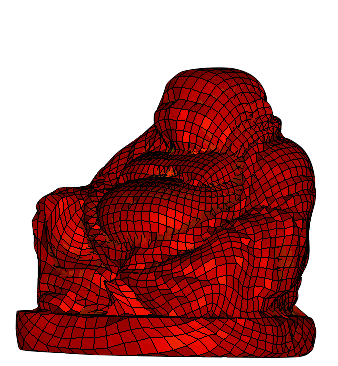}
      }}
    \mbox{ \centering
    \parbox{.45\columnwidth}{\centering
    (a)}
    \parbox{.45\columnwidth}{\centering
    (b)}}\\
\mbox{ \centering
\parbox{.45\columnwidth}{
\quad\quad \includegraphics[width=1.2in]{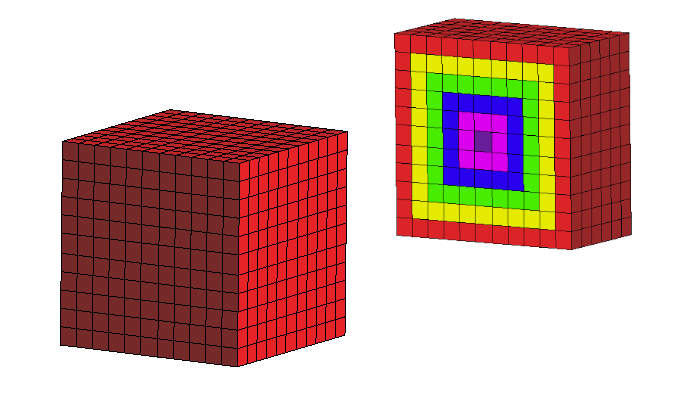}
      }\quad\quad\quad
\parbox{.45\columnwidth}{
    \includegraphics[width=1.2in]{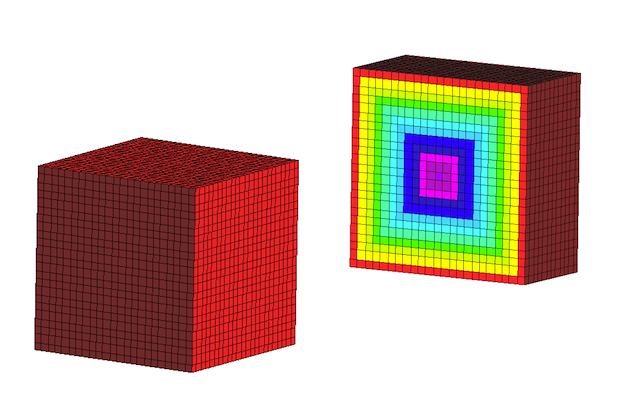}
      }}
    \mbox{ \centering
    \parbox{.45\columnwidth}{\centering
    (c)}
    \parbox{.45\columnwidth}{\centering
    (d)}}
\caption{An illustration of how the size of the solid cube impacts the level of details of the hexahedral mesh.
Top Row:  Two hexahedral meshes with differing levels of detail.  Bottom Row:  The corresponding solid cube and a mesh slice illustrate the resolution or level of detail.  Note: The color of each shell denotes the connectivity of model volume (not seen).}
\label{fig:Buddha_cubes}
\end{figure}
To construct $g$ we look for $g=g_1$ where $g_t$ is a one parameter
family of diffeomorphisms starting at $g_0=id_M$ (the identity map
on $M$) and evolving according to
\[
\frac{d g_t(x)}{dt} = X_t(g_t(x)).
\]
Let $\omega_t = (g_t)^*(\omega_1)$. We try to find vector fields $X_t$
such that $\omega_t=\omega_0 + t(\omega_1-\omega_0).$
This gives
\begin{eqnarray*}
\frac{d\omega_t}{dt} &=& \omega_1-\omega_0 \\
 &=& \frac{d(g_t)^*(\omega_1)}{dt} \\
 &=& L_{X_t}(\omega_t) \\
 &=& (d i_{X_t} + i_{X_t} d)(\omega_t) \\
 &=& (d i_{X_t})(\omega_t).
\end{eqnarray*}
Here $L_{X_t}$ denotes the Lie derivative, $i_{X_t}$ denotes the
interior product, and $d i_{X_t}$ the contraction; see \cite{Warner}
for details.  To get the vector fields $X_t$
we set the first and last lines equal to get
\[
(d i_{X_t})(\omega_t) = \omega_1-\omega_0
\]
Here the right hand side is an exact form (because the $\omega_i$ have
equal integrals and hence represent the same DeRahm cohomology class on $N$)
and hence there exists a one-form $\theta$ on $N$ with
$d\theta=\omega_1-\omega_0$. Finally, one can solve the equation
\[
(i_{X_t})(\omega_t) = \theta ,
\]
because the $\omega_t$ are volume forms, and hence nondegenerate.

So in order to construct the volume-preserving diffeomorphisms, we
find $\theta$ from $d\theta=\omega_1-\omega_0$, then solve the
above equation for $X_t$, and finally integrate the vector fields
$X_t$ to compute $g=g_1.$
\begin{figure*}
\begin{center}
\begin{tabular}{cccc}
\includegraphics[width=1in]{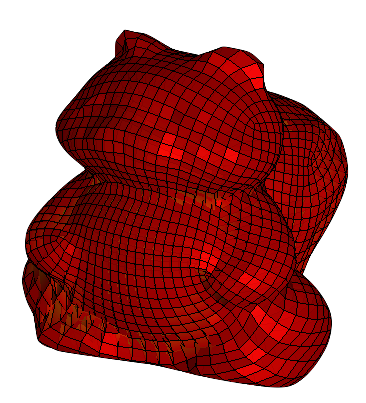}\quad\quad&
\includegraphics[width=1.15in]{buddha.png}\quad\quad&
\includegraphics[width=1in]{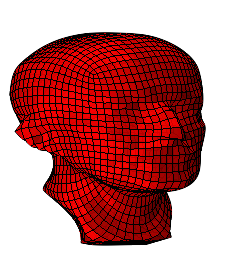}\quad\quad&
\includegraphics[width=.97in]{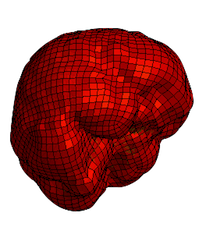}\\
\includegraphics[width=1in]{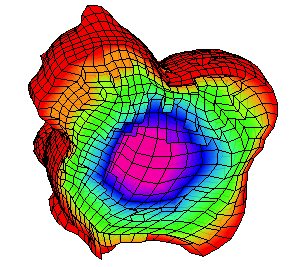}\quad\quad&
\includegraphics[width=1in]{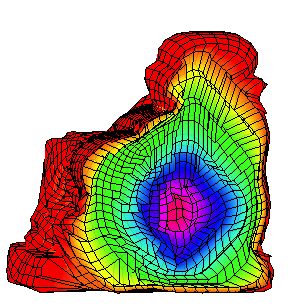}\quad\quad&
\includegraphics[width=.9in]{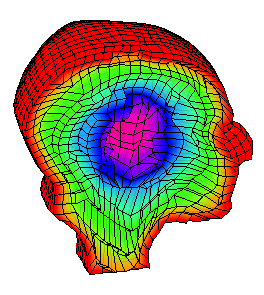}\quad\quad&
\includegraphics[width=1in]{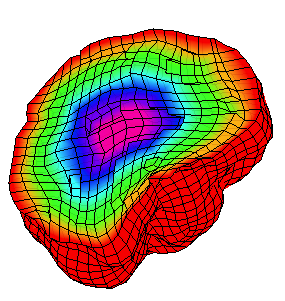}\\
(a) & (b) & (c) & (d)
\end{tabular}
\end{center}
\caption{Volumetric results are shown for a classical 3D models of varying complexity.  Specifically, the labeling is as follows:  (a) Squirrel. (b) Buddha. (c) Max Planck. (d). Brain. Top Row:  The surface of the hexahedral mesh.  Bottom Row:  A slice of mesh to illustrate the volumetric map where the shell color corresponds to a layer of a solid cube (not seen).}
\label{fig:several_models}
\end{figure*}
\subsubsection{Proposed Volume Preserving Flow}
In our case, one can explicitly write down these equations in  a form that makes them amenable to computer implementation. That is, we let $M$ be a solid which is diffeomorphic to the standard unit cube $C^3$ with the same total volume.  Moreover, $C^{3}$ corresponds to $N$ and the mapping $f : C^3\to M$ is the diffeomorphism obtained from Section~\ref{sec:initial_map}.
Then we want to find a one-parameter family of vector fields $X_t$ with $t\in [0,1]$ such that if we solve the ordinary differential equation \begin{equation}\label{Eq: Moser g from X}\frac{dg_t}{dt}= X_t \circ g_t\end{equation}
we get a family of diffeomorphisms $g_t$ : $C^3\to C^3$ with $g_0 = \mbox{identity}$, and
\[\vert\D g_t\vert((1 - t) \vert\D f\vert + t) = \vert\D f\vert\] where $\vert\D g_t\vert$ is the determinant of the Jacobian map $\D g_t$ and similarly for $ \vert\D f\vert$.
We want $g$ to have equal volume distribution, so in this case $\vert\D g_t\vert$ is $1$.
In order to find $X_t$, we solve
\begin{equation}\label{Eq: Moser v}\left\{ \begin{array}{ l l }
  \nabla\cdot v = 1-\vert\D f\vert  & \mbox{in }N  \\
  v = 0 & \mbox{on }\pd N  \\
\end{array} \right.\end{equation}
for $v$, and then \begin{equation}\label{Eq: Moser X from v} X_{t}=-\frac{v}{\left(1-t\right)\vert\D f\vert+t}.\end{equation}
to obtain our required flow. $\vert\D f\vert$ can be calculated as the ratio between corresponding volumes.
The required volume preserving diffeomorphism from $C^3$ to $M$ is $f \circ g(t=1)^{-1}$, where $f$ is the mapping obtained in Section \ref{sec:initial_map} and $g(t=1)$ is calculated using equation (\ref{Eq: Moser g from X}).  We then solve equation (\ref{Eq: Moser X from v}) and equation (\ref{Eq: Moser g from X}).  Note that as we evolve $g$ the location of our evaluation points changes, so in each step we need to interpolate the values of $\vert\D f\vert$ and $v$ at the new locations.  In the next section, we briefly discuss implementation and numerical details associated with the proposed algorithm.

\section{Numerical Implementation Details}
\label{sec:NumericDetailsImp}
In order to compute the above differential equations in discrete settings without changing their behavior, we use the discrete exterior calculus (DEC) machinery \cite{hirani03,desbrun06}.  The power of this method is the careful definition of discrete differential quantities, designed to respect structural relationships such as vector calculus identities.  This is quite different than previous methods, which focused on satisfying the continuous equations at a discrete set of spatial and temporal samples, but failed to preserve important global structures and invariants.  Thus, DEC provides tools in which one should store and manipulate quantities at their geometrically meaningful locations on the respective meshes (e.g., verticies, edges, faces, and volume elements).  We review the main concepts here, but refer the reader to \cite{hirani03,desbrun06} for complete details.

The main idea behind this approach is the representation of fields through measurements on cells: a 0-form represents a scalar function through its values at vertices (0-dim elements), while a 1-form represents a tangent vector field through its line integral along edges (1-dim elements).  This implies that tangent vector fields are specified as a single scalar per edge on the mesh.  A 2-form represents area density through its area integral over the faces (2-dim elements) and a 3-form represents volume density of a cell through its volume integral over cells (3-dim elements).  All relevant computations are then performed on these coefficients and the results are reconstructed with piecewise linear interpolation.

The theory itself defines discrete differential k-forms on meshes and express relevant operators such as divergence, curl, gradient, and Laplacian, as simple sparse matricies acting on intrinsic (coordinate-free) coefficients ``living'' on vertices, edges, faces, and cells.  To this end, these concepts were used and greatly simplified in redefining operators seen in the above equations.

Another note with regards to numerical implementation is the redundancy of the stereographic projection during the conformal mapping process.  Aside from a more aesthetic visualization, we opted for this projection as we had developed existing code from our previous work \cite{ayelet10} to perform interpolation and registration on the domain of a sphere.

Lastly, regarding the computational complexity, the proposed algorithm is limited by the conformal mapping process of the cube and volumetric object's confining surfaces.  This is the most computational costly step of the proposed algorithm as we have to map 2N surfaces to the sphere.  Unfortunately, we are not able to parallelize this portion of the algorithm because the extraction of the triangle in Section \ref{sec:initial_map} is chosen such that it is the closest distance, in the $L_2$ sense, to the adjacent layer or surface of higher resolution.  On the other hand, significant computational cost can be reduce if we incorporate coupled-active contours \cite{yezzi02}.  At the moment, the segmentation is done in a sequential manner for simplicity; however, the intent of future work is utilize several constrained active contours operating simultaneously to generate the subdivisions.  In turn, this should greatly improve the results, which are discussed next.
\section{Experiments and Evaluation}\label{sec:mesh_experiments}
We provide experimental results on our algorithm on simply connected meshes of varying complexity\footnote{All models, except the catalog of 3D prostates, were obtained and downloaded from the AIM@SHAPE Shape Repository: http://shapes.aim-at-shape.net/viewmodels.php} .  In particular, we focus on on both the quality of volume preserving map and the hexahedral mesh.  It should be noted that because code was not readily available, we do not claim the proposed method is superior (practically) to related techniques.  We also note that the following experiments were generated from unoptimized code written in MATLAB v.7.1 on a Intel Dual Core 2.66GHz with 4 GB memory.  Given that parts of the algorithm are parallelizable (e.g., GAC segmentation), we expect that computational times will increase significantly if one optimizes the code on the graphics processing unit (GPU) in conjunction with C++.  Thus, the experiments were performed to highlight the (dis)advantages of the proposed approach as an alternative to volumetric generation and hexahedral meshing algorithms.

\subsection{Qualitative Results:  Classic 3D Objects}
\label{sec:classic_qual}
\begin{table}[t]
\caption{Quantitative Results for Classic 3D Models} 
\centering 
\begin{tabular}{l l l l l  } 
\hline\hline 
Model & \# of     & \# of & vol & \% of   \\ [0.5ex] 
      &  nodes & hex. & var. &  concave    \\ [0.5ex] 
\hline 
Omotondo 1     & 1728  & 1331 & 0.14  & 9.50   \\
Omotondo 2     & 2744  & 2197 & 0.13  & 8.80    \\
Max-Planck 1   & 1728  & 1331 & 0.25  & 14.9   \\
Max-Planck 2   & 2744  & 2197 & 0.22  & 11.6   \\
squirrel       & 4096  & 3375 & 0.24  & 11.1   \\
tooth          & 1728  & 1331 & 0.18  & 16.4    \\
brain          & 2744  & 2197 & 0.13  & 7.20    \\
head           & 2744  & 2197 & 0.1   & 5.40   \\[1ex] 

\hline 
\end{tabular}
\label{table:performance} 
\end{table}
In the first set of the experiments, we provide qualitative results of our algorithm on several classical 3D models. Beginning with the Buddha model, we attempt to visually illustrate the output of our algorithm, with regards to level of mesh detail, for two different ``destination'' cubes. This can be seen in Figure~\ref{fig:Buddha_cubes}.  In particular, one can see that level of detail of the Buddha model associated with Figure~\ref{fig:Buddha_cubes}a to that of Figure~\ref{fig:Buddha_cubes}b is quite different.  This is due to the input parameter $N$ which dictates size of the constructed cube.  That is, in choosing $N$, we chose the number of shells to be used for both the input data and the cube as shown in Figure~\ref{fig:Buddha_cubes}c and Figure~\ref{fig:Buddha_cubes}d.  Initially shown in Figure 1, we note that each of the intermediate surfaces is assigned a given color that correlates to a shell on the corresponding cube. This color coding scheme will depict the connectivity, and will be used throughout the remainder of the experiments.
\begin{figure*}
\begin{center}
\begin{tabular}{ccccc}
\includegraphics[width=1in]{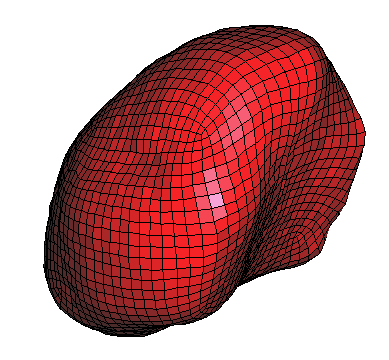}&
\includegraphics[width=1in]{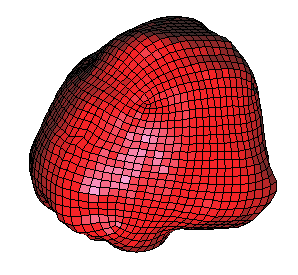}&
\includegraphics[width=1in]{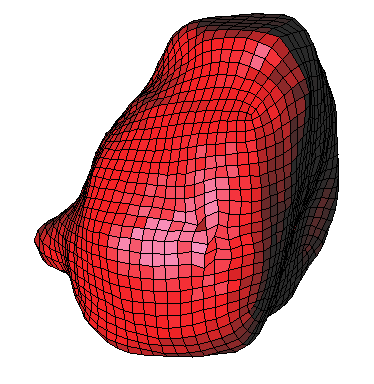}&
\includegraphics[width=1in]{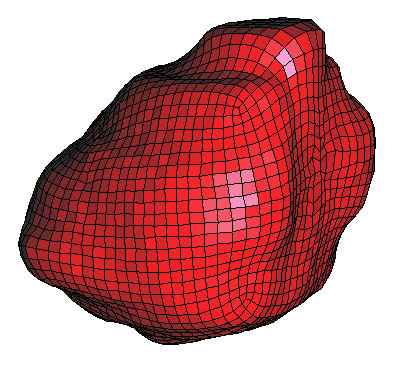}&
\includegraphics[width=1in]{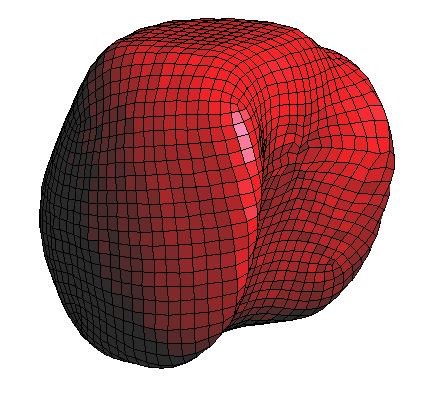}\\
\includegraphics[width=.9in]{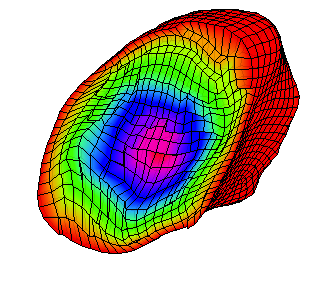}&
\includegraphics[width=1in]{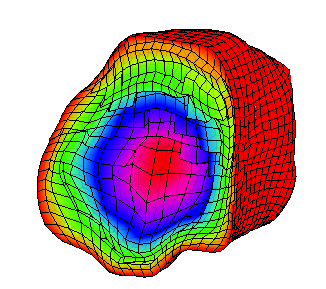}&
\includegraphics[width=1in]{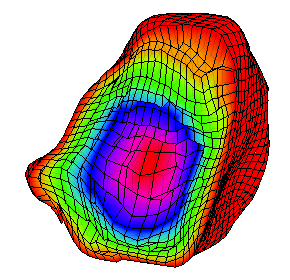}&
\includegraphics[width=1in]{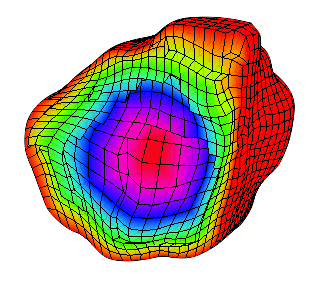}&
\includegraphics[width=1in]{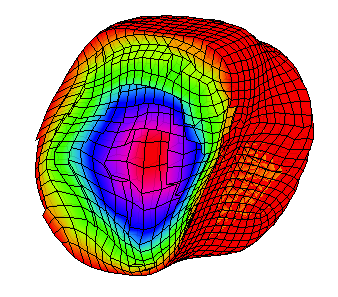}\\
(a) & (b) & (c) & (d) & (e)
\end{tabular}
\end{center}
\caption{Volumetric results are shown for a database of 3D prostates (5 of 10 shown) extracted from MRI imagery.  Specifically, the labeling is as follows:  (a) Prostate 1. (b) Prostate 2. (c) Prostate 3. (d). Prostate 4. (e) Prostate 10.  Top Row:  The surface of the hexahedral mesh.  Bottom Row:  Volumetric map shown via mesh slice where the shell color corresponds to a layer of a cube (not seen).}
\label{fig:prostate_cat}
\end{figure*}
 \begin{figure}[t]
 \begin{center}
 \includegraphics[width=2.5in]{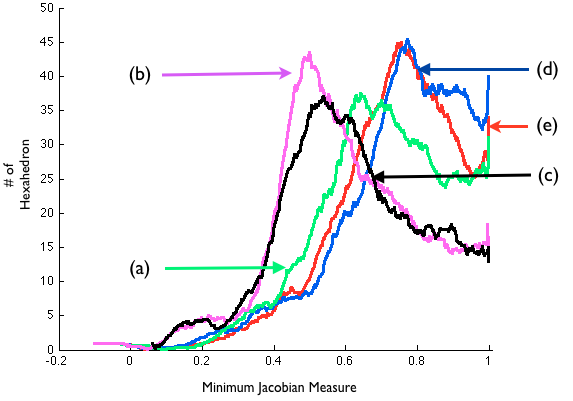}
 \end{center}
 \caption{Plot of the Minimum Jacobian Measure of meshes that correspond with Figure \ref{fig:several_models} and Figure \ref{fig:prostate_cat}.  Specifically, the labeling is as follows:  (a) Prostate 1. (b) Prostate 2. (c) Prostate 3. (d). Brain. (e) Max Planck.}
 \label{fig:minjacob}
 \end{figure}
\begin{table}[t]
\caption{Quantitative Results for Prostate Database} 
\centering 
\begin{tabular}{l l l l l l } 
\hline\hline 
Prostate & \!\!\!\!\! \hspace{8pt}\%  &vol&  \!\!\!\!\% concave  &  \!\!\!\!\hspace{7pt}vol. var.   &  \!\!\!\!\hspace{4pt}time.\\ [0.5ex] 
Model      & \!\!\!\!\! concave &var. & \!\!\!\! (w/smooth)   & \!\!\!\! (w/smooth) & \!\!\!\! (sec.)  \\ [0.5ex] 
\hline 
P. 1     & 6.52  & 0.33 &\hspace{3pt}  0.89  & 0.81  &  375  \\
P. 2    & 8.53  & 1.27 &\hspace{3pt}  1.21  & 0.34  &  310  \\
P. 3   & 6.62  & 1.11 &\hspace{3pt}  1.69  & 0.32 &  297   \\
P. 4  & 10.75  & 1.54 &\hspace{3pt}  0.54  & 0.41 &  393  \\
P. 5 & 4.75  & 1.28 & \hspace{3pt}  0.67  & 0.35  &  288   \\
P. 6  & 13.52  & 1.27 &\hspace{3pt}   3.19  & 0.55  &  284  \\
P. 7       & 18.37  & 1.54 &\hspace{3pt}   2.26  & 0.17 &  295  \\
P. 8         & 13.46  & 1.52 &\hspace{3pt}   1.02  & 0.15 &  302   \\
P. 9         & 11.52  & 0.86 &\hspace{3pt}   0.90  & 0.16  &  301  \\
P. 10           & 14.19  & 0.68 &\hspace{3pt}   1.24   & 0.11  &  392  \\[1ex] 

\hline 
\end{tabular}
\label{table:performance_2} 
\end{table}

With this being said, we then attempted to construct a volumetric preserving map for a 3D squirrel, the statue of Max Planck, and the cortex of a brain (see Figure \ref{fig:several_models}). Each of the models posses varying surface topology with concave regions that could hinder one's algorithm.  More importantly, while we have not shown the connectivity of the hexahedral mesh to that of the cube, the reader can rightfully infer that the given color code assigned to an intermediate surface does in fact correlate to a shell on a given cube.  Nevertheless, from a strictly visualization viewpoint, the proposed approach is able to output a quality hexahedral mesh with a seemingly good volumetric map.  Of course, the notion of of how we precisely define ``success'' will be the topic of the next section.
\vspace{-5pt}
\subsection{Quantitative Results:  Quality of Volumetric Map}
\vspace{-5pt}
In this section, we provide quantitative results on the aforementioned models in Section \ref{sec:classic_qual} as well as additional 3D models. Table~\ref{table:performance} summarizes some of our mapping results.  In general, after the initial map has been formed, hexahedra volume variance is found to be approximately $0.4$ where we note that average volume is normalized to be 1.   However, this value can fluctuate and depends usually on a specific structure. By improving the initial map, one can see that the final variance of the volume elements is reduced.

With regards to the quality of a hexahedral mesh, Table~\ref{table:performance} shows that models with concave areas may result in hexahedral meshes with concave elements. This may occur when portions of a concave region are mapped to a region near the edges of the solid cube, in which two faces of the model surface can be mapped to two faces of the same hexahedron. Figure~\ref{fig:concave_smooth} shows the concave elements in the brain model.  From this, one can see that some of concave elements are those that correspond to elements that appear on the edges of the solid cube. To overcome this degeneracy, smoothing the intermediate layers reduces the number of concave elements in the interior of the mesh, but does not eliminate them since these layers still follow the structure of the boundary of the given solid.

Thus, mesh smoothing can be applied to reduce the number of concave elements as seen in Figure~\ref{fig:concave_smooth}. A common smoothing strategy is based on the theory of harmonic functions and utilizes the discretized Laplacian, which enforces that the value of any point inside a region must be equal to the average of the values located about a neighborhood of the specified point.  In doing so, the smoothing operation may come at the expense of enlarging the variance of the volumetric elements.  Consequently, any smoothing that is incorporated is done so after the construction of the initial map.  Then any increase in volume variance will be negated when we seek to constrain our initial map to be volume preserving.  This is shown in the next section.
\subsection{Quantitative Results:  Medical Structures}
Although our method can be applied to general computer graphic models as seen above, one important application of our algorithm lies within medical imaging.  For example, longitudinal clinical studies involving imaging (e.g., tumor growth, surgical planning) is garnering much attention in the medical community as way to effectively understand the progress of a particular patient.  Thus, if we are able to generate a quality volumetric map, then the problem of multi-object shape analysis and atlas generation can be greatly simplified.
 \begin{figure}[t]
 \begin{center}
 \includegraphics[width=2.5in]{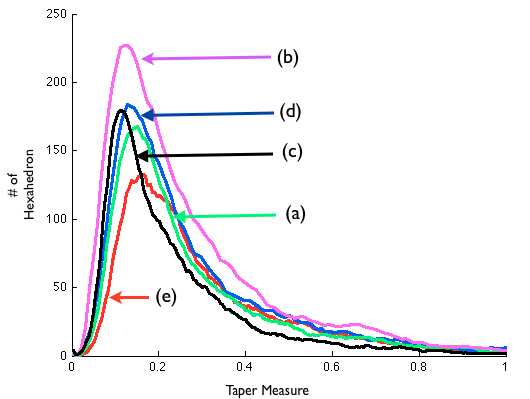}
 \end{center}
 \caption{Plot of the Taper Measure of meshes that correspond with Figure \ref{fig:several_models} and Figure \ref{fig:prostate_cat}.  Specifically, the labeling is as follows:  (a) Prostate 1. (b) Prostate 2. (c) Prostate 3. (d). Brain. (e) Max Planck}
 \label{fig:taper}
 \end{figure}

Using a catalog of 3D prostates, which were previously extracted offline, we construct the corresponding volumetric map and hexahedral mesh.  Specifically, we experimented with 10 shapes, five of which are shown in Figure~\ref{fig:prostate_cat}.  Using a destination cube with 8000 vertices (6859 hexahedron), we were able to obtain successful volumetric results.  As stated before, mesh smoothing may aid in the reduction of concave elements if ones employs it prior to evolving the mesh by the proposed volumetric flow.  To illustrate this, Table~\ref{table:performance_2} presents percentage of concave elements with and without smoothing. Note that the initial volumetric variance was typically above 10.  Interestingly, while we expect the number of concave elements to be smaller, the variance in the volumetric elements is reduced when smoothing is incorporated.  In explaining this phenomena, we should note that our volumetric flow is a variational method that is only locally convergent.  To this end, the initialization will indeed impact our result, and hence, lead to explaining small improvements in volume variance.

\begin{figure}[t]
 \begin{center}
 \includegraphics[width=2.5in]{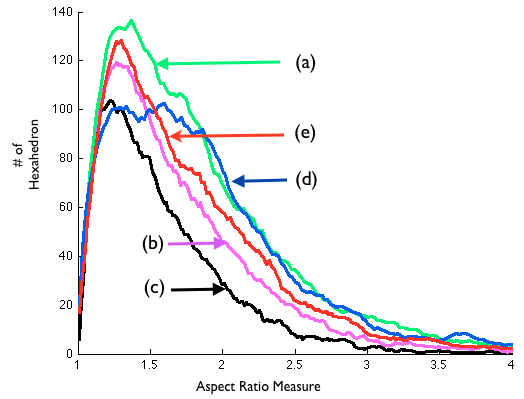}
 \end{center}
 \caption{Plot of the Aspect Ratio of meshes that correspond with Figure \ref{fig:several_models} and Figure \ref{fig:prostate_cat}.  Specifically, the labeling is as follows:  (a) Prostate 1. (b) Prostate 2. (c) Prostate 3. (d). Brain. (e) Max Planck.}
 \label{fig:aspect}
 \end{figure}

Moreover, to further validate the quality of the hexahedral mesh, we employ the commonly used minimum Jacobian, aspect ratio, and taper measure for five of the meshes shown in Figure \ref{fig:several_models} and Figure \ref{fig:prostate_cat}.  Specifically, Figure \ref{fig:minjacob} to Figure \ref{fig:aspect} plot the corresponding distribution for three prostates along with the Buddha and brain model.  With regards to the level of success or quality of mesh, the CUBIT project \cite{cubit} states that acceptable ranges for the minimum Jacobian, aspect ratio, and taper measure are [0.5 to 1], [1 to 4] , and [0 to 0.4], respectively.  The results of these measures for our generated meshes do indeed fall within the ranges stated. Moreover, while improvements can be made with post processing with the tradeoff of increasing volume variance, we believe that by deriving and incorporating a smoothing term in the second step of our algorithm, we can expect even higher quality meshes.  This will be a subject of future work.  Lastly, we point out that the computational time associated with the algorithm is reasonable, given that we are using unoptimized code.
\section{Conclusion}
\label{sec:mesh_future_work}
In this work, we presented a novel method to construct a volume preserving mapping from a cube to any given solid model.  By selecting a solid cube as our canonical domain of interest,  the process of volumetric meshing yields a regular hexahedral structure. To the best of our knowledge, this is the first attempt to create such a mapping.  Moreover, the method is robust and requires only the model and a resolution parameter as inputs.

Of course, with this baseline volumetric approach now developed, future work will concern itself with its applications to classical computer graphics and computer vision problems.  For example, we believe the volumetric map will enable us to rigorously treat the problem of multi-object shape analysis and atlas generation in a more robust fashion.  That is, using the proposed map, we will be able to transfer the problem of registration and analysis for several objects onto to the simplified cube.  In addition to the various possible applications, we intend to enhance the algorithm with the use of coupled active contours \cite{yezzi02} as well as employing the theory of optimal mass transport for volume preservation.  Like that of \cite{lipman2010}, we believe that incorporating theory from Monge Kantrovich will aid in the improvement of the proposed algorithm as a way to establish similarities between two interested objects.  In doing so, we should then be able to handle a more general class of 3D objects with higher precision.

{\small
\bibliographystyle{ieee}
\bibliography{template}
}
\end{document}